\documentclass{JHEP3}
\usepackage{graphics}

\title{On Graviphoton F-terms of $\mathcal{N}=1$ $SU(N)$ SYM with
Fundamental Matter }
\author{Ido Adam and Yaron Oz  \\
Raymond and Beverly Sackler Faculty of Exact Sciences \\
School of Physics and Astronomy \\
Tel-Aviv University, Ramat-Aviv 69978, Israel \\
E-mail: \email{adamido@post.tau.ac.il}, \email{yaronoz@post.tau.ac.il}}

\abstract{We consider $\mathcal{N}=1$ $SU(N_c)$ supersymmetric gauge
theory with chiral matter multiplets in the fundamental representation
of the gauge group. The general form of the meson
correlation functions in the presence of graviphoton background with
or without gravity is obtained. Finally, the  perturbation theory scheme
of computing these correlation functions is discussed.}

\preprint{hep-th/0506081\\TAUP-2805-05}


\newcommand{\ev}[1]{\ensuremath{\left\langle #1 \right\rangle}}
\newcommand{\tr}{\ensuremath{\mathrm{tr}}}
\newcommand{\Tr}{\ensuremath{\mathrm{Tr}}}

\begin{document}
\section{Introduction}
F-terms of four-dimensional supersymmetric gauge theories in
supergravity and graviphoton backgrounds have attracted much attention
in recent years.  On the one hand they are related to certain exactly
computable amplitudes of two gravitons and graviphotons.  On the other
hand they are computed by second quantized partition functions of
topological strings \cite{Antoniadis:1993ze}, and have an interesting
mathematical structure \cite{Bershadsky:1993cx}.  Gravitational
F-terms are directly related to the partition function of
two-dimensional non-critical strings
\cite{Dijkgraaf:2003xk,Ita:2004yn}.  Recently, gravitational F-terms
have been related to the computation of certain ${\cal N}=2$ black
hole partition functions \cite{Ooguri:2004zv}.

In this paper we will consider the gravitational and graviphoton
correlation functions in the context of four-dimensional ${\cal N}=1$
supersymmetric gauge theories.  Dijkgraaf and Vafa suggested a matrix
model description, where the gravitational F-terms can be computed by
summing up the non-planar matrix diagrams \cite{Dijkgraaf:2002dh}.
The assumption made is that the relevant fields are the glueball
superfields $S_i$ and the F-terms are holomorphic couplings of the
glueball superfields to gravity and the graviphoton.  The DV matrix
proposal has been proven diagrammatically in
\cite{Ooguri:2003qp,Ooguri:2003tt}.

The gravitational and graviphoton F-terms of interest are of the form
\begin{eqnarray}
\Gamma_1 & = & \sum_{g=0}^\infty \int d^4
x d^2 \theta (F_{\alpha \beta} F^{\alpha \beta})^g N_i \frac{\partial
F_g(S)}{\partial S_i} \ , \\
\Gamma_2 & = & \sum_{g=1}^\infty g \int
d^4 x d^2 \theta G^2 (F_{\alpha \beta} F^{\alpha \beta})^{g-1} F_g(S)
\ ,
\end{eqnarray}
where $G_{\alpha\beta\gamma}$ is the  ${\cal N}=1$ 
Weyl superfield and $F_{\alpha \beta}$ is the graviphoton.
According to the DV proposal, $F_g (S_i)$ is the partition function 
of the corresponding matrix model evaluated by
summing the genus $g$ diagrams with $S_i$ being the 't Hooft
parameters.

The approach we will take is to use the generalized Konishi anomaly
equations and some knowledge on the correlation functions to obtain
information about the perturbative contribution to the correlation
functions involving gravity and the graviphoton. In general, it is not
clear in which cases the generalized Konishi anomaly equations are
sufficient in order to determine the gravitational and graviphoton
F-terms. We will study $\mathcal{N}=1$ $SU(N_c)$ supersymmetric gauge
theory with fundamental matter and show that the anomaly equations are
insufficient to obtain the full perturbative correlation functions
(and the F-terms).  We will also discuss the field theoretic
perturbative diagrammatic computation.

Other recent works on the computation of gravitational and graviphoton
correlation functions and F-terms are
\cite{Klemm:2002pa,Dijkgraaf:2002yn,David:2003ke,Alday:2003ms,
Alday:2003dk,Gripaios:2003gw,Fuji:2004vf,Adam:2004sf}.

The paper is organized as follows. In section
\ref{sec:computational-scheme} we review the computational
scheme. Then in section \ref{sec:SUN-SYM-fund-matter} we apply the
scheme to the computation of correlation functions in $SU(N_c)$
SYM. Later we demonstrate the problems with the diagrammatic computation
in section \ref{sec:field-theory-graphs}.

\section{The computational scheme} \label{sec:computational-scheme}
Here we review the scheme used for computing the correlation functions
in the presence of either graviphoton or gravity backgrounds.

\subsection{The chiral ring}
We consider here an $\mathcal{N}=1$ supersymmetric gauge theory with
chiral matter multiplets coupled to it. Chiral operators are operators
annihilated by the covariant derivative $\bar D_{\dot \alpha}$. All
such operators modulo terms which are $\bar D_{\dot \alpha}$ exact
form the ring structure of the chiral ring.

Denoting by $W_\alpha=-\frac{1}{4}\bar D^2 e^{-V} D_\alpha e^V$ the
spinor field-strength superfield of the vector superfield $V$ one has
in flat space that in the chiral ring (i.e., up to $\bar D_{\dot
\alpha}$ exact terms)
\begin{equation}
\{W_\alpha, W_\beta\} = 0 \ . \label{eq:undeformed-anticommutator}
\end{equation}
This relation is modified in a background of gravity and the
graviphoton field. Let $G_{\alpha \beta \gamma}$ be the
$\mathcal{N}=1$ Weyl superfield and $F_{\alpha\beta}$ be the
graviphoton field which together form the $\mathcal{N}=2$ Weyl
superfield $H_{\alpha\beta}=F_{\alpha\beta}+\hat\theta^\gamma
G_{\alpha\beta\gamma}$, where $\hat\theta$ is the additional
supercoordinate of $\mathcal{N}=2$ superspace. In the presence of
these either the supercoordinates become non-anti-commutative or
(\ref{eq:undeformed-anticommutator}) is modified
\cite{Ooguri:2003qp,Ooguri:2003tt,deBoer:2003dn} to
\begin{equation}
\{W_\alpha, W_\beta \} = F_{\alpha \beta} + 2 G_{\alpha \beta \gamma}
W^\gamma \ . \label{eq:C-deformation}
\end{equation}
This deformation of the chiral ring leads to the chiral ring relations
\cite{Alday:2003ms}
\begin{eqnarray}
[W_\alpha, W^2] & = & -2 F_{\alpha \beta} W^\beta \ ,
\label{eq:chiral-ring1} \\
\{W_\alpha, W^2\} & = & -\frac{2}{3} (G^2 W_\alpha + G_{\alpha \beta
\gamma} F^{\beta \gamma}) \ , \\
W_\alpha W^2 & = & -F_{\alpha \beta} W^\beta - \frac{1}{3} G^2
W_\alpha - \frac{1}{3} G_{\alpha \beta \gamma} F^{\beta \gamma} \ ,
\label{eq:chiral-ring3} \\
(G^2)^2 & = & 0 \ .
\end{eqnarray}

In addition, for a chiral superfield in the fundamental or
anti-fundamental representation \cite{David:2003dh}
\begin{equation}
W_{\alpha a}{}^b Q_b^i = \tilde Q^a_i W_{\alpha a}{}^b = 0
\label{eq:chiral-ring6}
\end{equation}
in the chiral ring.

\subsection{The Konishi anomaly equations}
The classical Konishi equations in the chiral ring for a field
transformation $\delta Q_a^i$ are
\begin{equation}
\frac{\partial W_\mathrm{tree}}{\partial Q_a^i} \delta Q_a^i = 0 \ .
\end{equation}
These are modified quantum mechanically
\cite{Konishi:1985tu,Konishi:1984hf} to
\begin{equation}
\ev{\frac{\partial W_\mathrm{tree}}{\partial Q_a^i} \delta Q_a^i} +
\ev{\left( \frac{1}{32 \pi^2} W_{\alpha a}{}^b W^\alpha{}_b{}^c +
\frac{1}{32 \pi^2} \frac{G^2}{3} \delta^c_a \right) \frac{\partial
\delta Q_c^i}{\partial Q_a^i}} = 0 \ . \label{eq:gen-grav-Konishi}
\end{equation}
As argued in \cite{Alday:2003ms}, the Konishi anomaly equations are
not modified in the presence of the graviphoton field. The argument is
based on the graviphoton being of dimension three so any Lorentz
scalar with smooth limits of the dimensional parameters of the theory
constructed from it would have a dimension greater then three. All the
terms in the Konishi anomaly equation are of dimension three,
therefore the anomaly equation for a field transformation $\delta
Q_a^i$ remains of the form (\ref{eq:gen-grav-Konishi}) and is not
modified.

In the next section we will use the Konishi anomaly equations in order
to obtain perturbative information on the correlation functions. As
will be seen, the Konishi equations are not sufficient to completely
determine the correlations functions, but they do provide some
constraints on their general form.

\section{$SU(N_c)$ SYM with fundamental matter} \label{sec:SUN-SYM-fund-matter}
In this section we consider the $\mathcal{N}=1$ $SU(N_c)$ SYM theory
with chiral matter multiplets $Q_a^i$ and $\tilde Q^a_i$
($a,b,\dots=1,\dots,N_c$ are color indices and $i,
j,\dots=1,\dots,N_f$ are flavor indices) in the fundamental and
anti-fundamental representation, respectively, considered in
\cite{Brandhuber:2003va,Ita:2003kk}. The theory has the tree-level
superpotential
\begin{equation}
W_\mathrm{tree} = m \tr M + \lambda \tr M^2 \ ,
\end{equation}
where $M_i{}^j = \tilde Q^a_i Q_a^j$ are the gauge-invariant meson
operators. For simplicity we will take the case of a single flavor
($N_f=1$), but the results should be readily extendable to $N_f>1$.

\subsection{The anomaly equations}
We first look at the field transformation $\delta Q_a=Q_a M^k$. It
yields the equations obtained in \cite{Ita:2003kk}
\begin{equation}
m \ev{M^{k + 1}} + 2 \lambda \ev{M^{k + 2}} + \frac{N_c + k}{3} G^2
\ev{M^k} = S \ev{M^k} \ . \label{eq:grav-Konishi-eq}
\end{equation}
(Here and henceforth we redefine the Weyl superfield $G^2\to G^2/32
\pi^2$.)

Since the graviphoton does not appear explicitly in the anomaly
equation, the only way to obtain graviphoton dependence is via the
chiral ring relations
(\ref{eq:chiral-ring1})--(\ref{eq:chiral-ring3}) or by including it
in the transformation $\delta Q_a$. However, using Lorentz invariant
graviphoton terms such as $F_{\alpha\beta} F^{\alpha\beta}$ in the
transformation will only result in multiplying the entire anomaly
equation by these expressions and not yield any new independent
equations.

$\delta Q$ should be a scalar in the fundamental representation of the
gauge group, so $W_\alpha$ can be incorporated into it only as
$W_{\alpha a}{}^b W^\alpha{}_b{}^c Q_c$, which vanishes in the chiral
ring, or as $S=-\frac{1}{32\pi^2}\Tr(W_\alpha W^\alpha)$ which will only
multiply the equation by $S$. Another possible scalar can be
constructed using the graviphoton: $F_{\alpha\beta} W^\alpha
W^\beta$. By using the symmetry of $F_{\alpha\beta}$ and
(\ref{eq:C-deformation}) one has
\begin{equation}
F_{\alpha \beta} W^\alpha W^\beta = \frac{1}{2} F_{\alpha \beta}
F^{\alpha \beta} + F_{\alpha \beta} G^{\alpha \beta}{}_\gamma W^\gamma
\ .
\end{equation}
This is in the adjoint representation so an appropriate term can be
obtained by acting with it on $Q$ which vanishes in the chiral ring
due to (\ref{eq:chiral-ring6}) or by tracing over the gauge indices,
making the second term vanish as $W_\alpha$ is traceless.

In general, transformations yielding the graviphoton will either vanish
because of (\ref{eq:chiral-ring6}) or will lead to dependent anomaly
equations. This is unlike the case of the theory with matter in the
adjoint representation of the gauge group since in that case
transformations such as $W_\alpha \Phi$ do not vanish and can combine
with the $W^2$ term in the anomaly equation to generate coupling to
the graviphoton by the chiral ring relations.

\subsection{Correlation functions without gravity}
\subsubsection{Solution of the anomaly equations}
The generalized Konishi anomaly equations for the theory considered
were found in the presence of gravity but no graviphoton backgrounds
in \cite{Ita:2003kk} and with gravity turned off are of the form
\begin{equation} \label{eq:generalized-Konishi}
S \ev{M^k} = m \ev{M^{k+1}} + 2 \lambda \ev{M^{k+2}} \ .
\end{equation}
Since the graviphoton background does not modify the anomaly equations
\cite{Alday:2003ms}, these remain valid even with the graviphoton
turned on.

By performing a $z$-transform of (\ref{eq:generalized-Konishi}) one
obtains the single equation
\begin{equation}
S \sum_{k=0}^\infty \frac{1}{k!} \ev{M^k} z^k = m
\sum_{k=0}^\infty \frac{1}{k!} \ev{M^{k+1}} z^k + 2 \lambda
\sum_{k=0}^\infty \frac{1}{k!} \ev{M^{k+2}} z^k \ .
\end{equation}
We now define the generating function for the meson operator
correlation functions
\begin{equation} \label{eq:gen-function}
f(z) = \sum_{k=0}^\infty \frac{1}{k!} \ev{M^k} z^k \ .
\end{equation}
It follows immediately that
\begin{eqnarray}
\frac{df(z)}{dz} & = & \sum_{k=0}^\infty \frac{1}{k!} \ev{M^{k+1}} z^k
\ , \label{eq:gen-function-1deriv}\\
\frac{d^2 f(z)}{dz^2} & = & \sum_{k=0}^\infty \frac{1}{k!}
\ev{M^{k+2}} z^k \ . \label{eq:gen-function-2deriv}
\end{eqnarray}
Hence, the infinite set of equations (\ref{eq:generalized-Konishi})
can be written as the ordinary differential equation
\begin{equation}
S f(z) = m \frac{df(z)}{dz} + 2 \lambda \frac{d^2 f(z)}{dz^2} \ ,
\end{equation}
whose general solution is
\begin{equation} \label{eq:no-grav-f-solution}
f(z) = A_+ \exp\left( \frac{-m + \sqrt{m^2 + 8 \lambda S}}{4 \lambda}
z \right) + A_- \exp\left( \frac{-m - \sqrt{m^2 + 8 \lambda S}}{4 \lambda}
z \right) \ ,
\end{equation}
where $A_+$ and $A_-$ are coefficients which may depend on the
couplings as well as the glueball superfield $S$ and the
graviphoton. Particularly, note that any possible graviphoton
dependence may enter through these coefficients alone.

Thus, we conclude that the correlation functions are of the form
\begin{equation}
\ev{M^k} = A_+ \left( \frac{-m + \sqrt{m^2 + 8 \lambda S}}{4 \lambda}
\right)^k + A_- \left( \frac{-m - \sqrt{m^2 + 8 \lambda S}}{4 \lambda}
\right)^k \ .
\end{equation}
Because correlation functions of chiral operators factorize in the
absence of the graviphoton and gravity, in the limit $F_{\alpha \beta}
\to 0$ either $A_+ \to 1$ and $A_- \to 0$ or the other way around
depending on whether the Higgsed vacuum is considered or not. Hence,
in the un-Higgsed vacuum
\begin{equation}
A_+ = 1 + O(F_{\alpha \beta} F^{\alpha \beta}) \ , \quad A_- =
O(F_{\alpha \beta} F^{\alpha \beta})
\end{equation}
and 
\begin{equation}
A_+ = O(F_{\alpha \beta} F^{\alpha \beta}) \ , \quad A_- = 1 +
O(F_{\alpha \beta} F^{\alpha \beta})
\end{equation}
in the Higgsed vacuum.

\subsubsection{The form of $A_\pm$}
As noted before, the graviphoton dependence can only enter via the
coefficients $A_\pm$. In general, these should depend on the couplings
$m$ and $\lambda$ and on the background fields $S$ and
$F_{\alpha\beta}$. Hence, the $A_\pm$ should be sums of terms of the
form
\begin{displaymath}
\lambda^n m^p S^q (F_{\alpha \beta} F^{\alpha \beta})^r \ .
\end{displaymath}
From holomorphicity we expect $n$, $p$, $q$ and $r$ to be integers. Also
since the coefficients are dimensionless, the powers must satisfy
\begin{equation}
-n + p + 3 q + 6 r = 0 \ .
\end{equation}
The limit $F_{\alpha \beta}\to 0$ must be regular so that the ordinary
correlation functions without graviphoton background obtained in
\cite{Brandhuber:2003va} are recovered. Thus $r\ge0$.  The classical
limit $S\to0$, in which the Konishi anomaly vanishes, must be smooth
so $q\ge0$. In the limit of $\lambda\to0$ the correlation function
\ev{M} is either smooth for the case of the un-Higgsed vacuum (the one
corresponding to the plus sign solution) or diverges as $1/\lambda$ in
the Higgsed vacuum (the minus sign vacuum). Thus, we do not expect
additional, higher order divergence as $\lambda\to0$ so $n\ge0$.

Finally, the flavor symmetries $U(1)_Q$ and $U(1)_{\tilde Q}$ are broken
at tree-level by the superpotential. These can be restored by
assigning charges to the couplings as given in the table
\begin{center}
\begin{tabular}{c|cc}
& $U(1)_Q$ & $U(1)_{\tilde Q}$ \\
\hline
$Q$ & $1$ & $0$ \\
$\tilde Q$ & $0$ & $1$ \\
$S$ & $0$ & $0$ \\
$m$ & $-1$ & $-1$ \\
$\lambda$ & $-2$ & $-2$ \\
$F_{\alpha \beta}$ & $0$ & $0$
\end{tabular}
\end{center}
Requiring the correlation functions to be invariant under these
restored symmetries yields
\begin{equation}
2 n + p = 0 \ .
\end{equation}
Putting all of this together we have
\begin{equation}
n = q + 2 r
\end{equation}
and the terms in the power series expansion of $A_\pm$ are of the form
\begin{equation}
\left( \frac{\lambda}{m^2} \right)^{q + 2 r} (F_{\alpha \beta}
F^{\alpha \beta})^r S^q \ .
\end{equation}
It should be noted that only terms which vanish in the limit
$m\to\infty$ are allowed --- in accordance with one's expectation that
the matter completely decouples in this limit, leaving only the pure
gauge theory coupled to gravity and to the graviphoton.

\subsection{Correlation functions with gravity and graviphoton
backgrounds}
\subsubsection{Solution of the anomaly equations}
By performing a $z$-transform on (\ref{eq:grav-Konishi-eq}) the following
equation is obtained
\begin{eqnarray}
S \sum_{k=0}^\infty \frac{z^k}{k!} \ev{M^k} & = & m \sum_{k=0}^\infty
\frac{z^k}{k!} \ev{M^{k + 1}} + 2 \lambda \sum_{k=0}^\infty
\frac{z^k}{k!} \ev{M^{k+2}} + \frac{N_c}{3} G^2 \sum_{k=0}^\infty
\frac{z^k}{k!} \ev{M^k} + \nonumber \\ 
&& {} + \frac{1}{3} G^2 \sum_{k=0}^\infty \frac{k z^k}{k!} \ev{M^k} \ .
\end{eqnarray}
Using the generating function (\ref{eq:gen-function}) one finds that
\begin{equation}
z \frac{df(z)}{dz} = \sum_{k=0}^\infty \frac{k z^k}{k!} \ev{M^k}
\end{equation}
 and together with the relations (\ref{eq:gen-function-1deriv}) and
(\ref{eq:gen-function-2deriv}) (\ref{eq:grav-Konishi-eq}) can be cast
into the ordinary differential equation
\begin{equation} \label{eq:grav-konishi-ode}
S f(z) = m \frac{df(z)}{dz} + 2 \lambda \frac{d^2 f(z)}{dz^2} +
\frac{N_c}{3} G^2 f(z) + \frac{1}{3} G^2 z \frac{df(z)}{dz} \ . 
\end{equation}

The function $f(z)$ can be expanded in powers of $G^2$ using the
chiral ring relation $(G^2)^2=0$ \cite{Alday:2003ms,David:2003ke} 
\begin{equation}
f(z) = f_0 (z) + G^2 f_1 (z)
\end{equation}
and upon substitution in (\ref{eq:grav-konishi-ode}) we get two
differential equations for $f_0(z)$ and $f_1(z)$:
\begin{eqnarray}
S f_0(z) & = & m \frac{df_0(z)}{dz} + 2 \lambda \frac{d^2
f_0(z)}{dz^2} \ , \\
S f_1(z) & = & m \frac{df_1(z)}{dz} + 2 \lambda \frac{d^2
f_1(z)}{dz^2} +\frac{N_c}{3} f_0(z) + \frac{1}{3} z \frac{df_0(z)}{dz}
\ . \label{eq:f1-ode}
\end{eqnarray}
Defining
\begin{equation}
\alpha_\pm = \frac{-m \pm \sqrt{m^2 + 8 \lambda S}}{4 \lambda} \ ,
\end{equation}
the solution of the equation for $f_0(z)$ is
(\ref{eq:no-grav-f-solution})
\begin{equation}
f_0(z) = A_+ e^{\alpha_+ z} + A_- e^{\alpha_- z} \ .
\end{equation}
Plugging this in (\ref{eq:f1-ode}) the following equation is obtained
\begin{eqnarray}
S f_1 & = & m \frac{df_1}{dz} + 2 \lambda \frac{d^2 f_1}{dz^2} +
\frac{N_c}{3} \left( A_+ e^{\alpha_+ z} + A_- e^{\alpha_- z} \right) +
\nonumber \\
&& {} + \frac{1}{3} z \left( \alpha_+ A_+ e^{\alpha_+ z} + \alpha_-
A_- e^{\alpha_- z} \right) \ .
\end{eqnarray}
By considering a solution of the form
\begin{equation}
f_1(z) = c_+(z) e^{\alpha_+ z} + c_-(z) e^{\alpha_- z}
\end{equation}
and further assuming that the equation thus obtained can be divided
into separate equations for the unknown functions $c_\pm(z)$ we have
\begin{equation}
2 \lambda \frac{d^2 c_\pm}{dz^2} + (m + 4 \lambda \alpha_\pm) \frac{d
c_\pm}{dz} + (m \alpha_\pm + 2 \lambda \alpha_\pm^2 -S) c_\pm +
\frac{N_c}{3} A_\pm + \frac{1}{3} z \alpha_\pm A_\pm = 0 \ ,
\end{equation}
whose solution is given by
\begin{eqnarray}
c_\pm(z) & = & \pm \frac{C_{1 \pm}}{\sqrt{m^2 + 8 \lambda S}} - \frac{N_c}{3}
A_\pm \left( \pm \frac{z}{\sqrt{m^2 + 8 \lambda S}} - \frac{2
\lambda}{m^2 + 8 \lambda S} \right) - \nonumber \\
&& {} - \frac{1}{6} \alpha_\pm A_\pm \left[ \pm \frac{8
\lambda^2}{(m^2 + 8 \lambda S)^{3/2}} - \frac{4 \lambda z}{m^2 + 8
\lambda S} \pm \frac{z^2}{\sqrt{m^2 + 8 \lambda S}} \right] +
\nonumber \\
&& {} + C_{2 \pm} \exp\left (\mp \frac{\sqrt{m^2 + 8 \lambda S}}{2 \lambda}
z \right) \ ,
\end{eqnarray}
where $C_{1\pm}$ and $C_{2\pm}$ are integration constants which may
depend on the couplings, the glueball superfield and the
graviphoton.

\subsubsection{Constraints on the coefficients}
As shown in \cite{Ooguri:2003tt}, there are two types of related
effective F-terms coupling the glueball to gravity and the graviphoton
\begin{eqnarray}
\Gamma_1 & = & \int d^4 x d^2 \theta W_0 = \sum_{g=0}^\infty \int d^4
x d^2 \theta (F_{\alpha \beta} F^{\alpha \beta})^g N_i \frac{\partial
F_g(S)}{\partial S_i} \ , \\
\Gamma_2 & = & \int d^4 x d^2 \theta G^2 W_1 = \sum_{g=1}^\infty g \int
d^4 x d^2 \theta G^2 (F_{\alpha \beta} F^{\alpha \beta})^{g-1} F_g(S)
\ .
\end{eqnarray}
Since the function $F_g(S)$ is found in both, in the case of unbroken
gauge symmetry the two are related as
\begin{equation}
\frac{\partial W_0}{\partial u} = N_c \frac{\partial W_1}{\partial S}
\ ,
\end{equation}
where we have set $u=F_{\alpha \beta} F^{\alpha \beta}$.

The correlation functions \ev{M} and \ev{M^2} can be obtained from the
effective superpotential $W_\mathrm{eff}=W_0+W_1 G^2$ by
differentiating it with respect to the couplings,
\begin{equation}
\ev{M} = \frac{\partial W_\mathrm{eff}}{\partial m} \ , \quad \ev{M^2} =
\frac{\partial W_\mathrm{eff}}{\partial \lambda} \ .
\end{equation}
Hence, $f_0(z)$ and $f_1(z)$ must satisfy the relations
\begin{eqnarray}
\left. \frac{\partial^2 f_0}{\partial u \partial z} \right|_{z=0} & =
& N_c \left .\frac{\partial^2 f_1}{\partial S \partial z} \right|_{z=0}
\ , \\
\left. \frac{\partial^3 f_0}{\partial u \partial z^2} \right|_{z=0} & =
& N_c \left .\frac{\partial^3 f_1}{\partial S \partial z^2}
\right|_{z=0} \ .
\end{eqnarray}
The constraint on the integration constants from the first of these is
\begin{eqnarray}
0 & = & \frac{2 N_c \lambda \left[ (1+N_c) m^2 + 4 (2 N_c - 1) \lambda
S - 2 N_c m \sqrt{m^2 + 8 \lambda S} \right]}{3 (m^2 + 8 \lambda
S)^{5/2}} A_- - \nonumber \\
&& {} - \frac{2 N_c \lambda \left[ (1+N_c) m^2 + 4 (2 N_c - 1) \lambda
S + 2 N_c m \sqrt{m^2 + 8 \lambda S} \right]}{3 (m^2 + 8 \lambda 
S)^{5/2}} A_+ + \nonumber \\
&& {} + \frac{N_c m (C_{1-} - C_{1+})}{(m^2 + 8 \lambda S)^{3/2}} +
\frac{N_c (C_{2+} - C_{2-})}{\sqrt{m^2 + 8 \lambda S}} + \nonumber \\
&& {} + \frac{-m + \sqrt{m^2 + 8 \lambda S}}{4 \lambda} \frac{\partial
A_+}{\partial u} + \frac{-m - \sqrt{m^2 + 8 \lambda S}}{4 \lambda}
\frac{\partial A_-}{\partial u} + \nonumber \\
&& {} +\frac{N_c \left[ -4 \lambda S + N_c \left( m^2 + 8 \lambda S + m
\sqrt{m^2 + 8 \lambda S} \right) \right]}{6 (m^2 + 8 \lambda S)^{3/2}}
 \frac{\partial A_+}{\partial S} + \nonumber \\
&& {} + \frac{N_c \left[ 4 \lambda S - N_c \left( m^2 + 8
\lambda S - m \sqrt{m^2 + 8 \lambda S} \right) \right]}{6 (m^2 + 8
\lambda S)^{3/2}} \frac{\partial A_-}{\partial S} + \nonumber \\
&& {} + \frac{N_c (m - \sqrt{m^2 + 8 \lambda S})}{4 \lambda \sqrt{m^2 +
8 \lambda S}} \frac{\partial C_{1+}}{\partial S} - \frac{N_c (m
+ \sqrt{m^2 + 8 \lambda S})}{4 \lambda \sqrt{m^2 + 8 \lambda S}}
\frac{\partial C_{1-}}{\partial S} + \nonumber \\
&& {} + \frac{N_c (m + \sqrt{m^2 + 8 \lambda S})}{4 \lambda} \frac{\partial
C_{2+}}{\partial S} + \frac{N_c (m - \sqrt{m^2 + 8 \lambda S})}{4
\lambda} \frac{\partial C_{2-}}{\partial S} \ ,
\end{eqnarray}
while the second one yields
\begin{eqnarray}
\lefteqn{0 = \frac{N_c m \left[ (1 + 2 N_c) m (-m + \sqrt{m^2 + 8
\lambda S}) + 4 (1 - 4 N_c) \lambda S \right] A_-}{6 (m^2 + 8 \lambda
S)^{5/2}} +} \nonumber \\
&& {} + \frac{N_c m \left[ (1 + 2 N_c) m (m + \sqrt{m^2 + 8 \lambda S}) - 4
(1 - 4 N_c) \lambda S \right] A_+}{6 (m^2 + 8 \lambda S)^{5/2}} +
\nonumber \\
&& {} + \frac{2 N_c S (C_{1-} - C_{1+})}{(m^2 + 8 \lambda S)^{3/2}} +
\frac{N_c (m - \sqrt{m^2 + 8 \lambda S}) C_{2-}}{2 \lambda \sqrt{m^2 +
8 \lambda S}} - \frac{N_c (m + \sqrt{m^2 + 8 \lambda S}) C_{2+}}{2
\lambda \sqrt{m^2 + 8 \lambda S}} + \nonumber \\
&& {} + \frac{(m + \sqrt{m^2 + 8 \lambda S})^2}{16 \lambda^2}
\frac{\partial A_-}{\partial u} + \frac{(-m + \sqrt{m^2 + 8 \lambda
S})^2}{16 \lambda^2} \frac{\partial A_+}{\partial u} + \nonumber \\
&& {} + \frac{N_c \left[ N_c m^2 (m + \sqrt{m^2 + 8 \lambda S}) + 2 (4
N_c -1) m \lambda S + 2 (6 N_c +1) \lambda S \sqrt{m^2 + 8 \lambda S}
\right]}{12 \lambda (m^2 + 8 \lambda S)^{3/2}} \frac{\partial
A_-}{\partial S} - \nonumber \\
&& {} - \frac{N_c \left[ N_c m^2 (m - \sqrt{m^2 + 8 \lambda S}) + 2 (4
N_c -1) m \lambda S - 2 (6 N_c +1) \lambda S \sqrt{m^2 + 8 \lambda S}
\right]}{12 \lambda (m^2 + 8 \lambda S)^{3/2}} \frac{\partial
A_+}{\partial S} + \nonumber \\
&& {} + \frac{N_c (m + \sqrt{m^2 + 8 \lambda S})^2}{16 \lambda^2
\sqrt{m^2 + 8 \lambda S}} \frac{\partial C_{1-}}{\partial S} -
\frac{N_c (-m + \sqrt{m^2 + 8 \lambda S})^2}{16 \lambda^2 \sqrt{m^2 + 8
\lambda S}} \frac{\partial C_{1+}}{\partial S} - \nonumber \\
&& {} - \frac{N_c (-m + \sqrt{m^2 + 8 \lambda S})^2}{16 \lambda^2}
\frac{\partial C_{2-}}{\partial S} - \frac{N_c (m + \sqrt{m^2 + 8
\lambda S})^2}{16 \lambda^2} \frac{\partial C_{2+}}{\partial S} \ .
\end{eqnarray}

\section{The field theory graphs} \label{sec:field-theory-graphs}
The entire Lagrangian of the $C$-deformed field theory is not
known. Hence it is not clear how to compute the correlation functions
in presence of graviphoton background. In this section we make some
assumptions about the Lagrangian and demonstrate the difficulties
arising from these assumptions.

According to the Dijkgraaf--Vafa conjecture the perturbative expansion
of the graviphoton correction terms should be obtainable by computing
the non-planar graphs of the theory. In
\cite{Ooguri:2003qp,Ooguri:2003tt} a scheme was given for computing
such graphs for matter in the adjoint representation of the gauge
group. This scheme does not appear to be applicable in the case of
matter in the fundamental representation since the scheme includes the
selection rule that a non-vanishing graph with $h$ holes should have
gaugino insertions in $h-1$ of its holes. This selection rule originated
from the requirement that the graphs be path-ordering
independent. Basically, for a graph to be path-ordering independent
one must have
\begin{equation}
\oint_{\gamma_i} p_\alpha = 0
\end{equation}
in that graph, where $p_\alpha$ is the world-sheet current of
space-time supersymmetry and $\gamma_i$ is the contour of the $i$-th
hole. This was enforced by inserting $h-1$ gaugino insertions
\begin{displaymath}
\prod_{i=1}^{h - 1} \left( \oint_{\gamma_i} W^\alpha p_\alpha \right)^2
\end{displaymath}
and utilizing the fact that
\begin{equation} \label{eq:susy_conservation}
\sum_{i=1}^h \oint_{\gamma_i} p_\alpha = 0 \ .
\end{equation}
However, (\ref{eq:susy_conservation}) does not hold in the case of
matter in the fundamental representation. This is most easily seen in
the field theory limit, in which (\ref{eq:susy_conservation}) takes
the form \cite{Ooguri:2003qp}
\begin{equation} \label{eq:field-theory-limit}
\sum_{i = 1}^h \sum_I s_I L_{I i} \pi_\alpha^I = 0 \ ,
\end{equation}
where the index $I$ denotes the propagator and $L_{I i}$ is a matrix
relating index-loop momenta to the propagator momenta.  Unlike the
double-line propagators of matter in the adjoint representation,
propagators of matter in the fundamental representation have only a
single color line. Therefore, whereas in the adjoint case each
propagator is traversed twice in opposite directions and thus
contributes two identical terms with opposite signs to the sum, so
(\ref{eq:field-theory-limit}) is satisfied, in the fundamental
case it is traversed once and obviously the sum can no longer vanish.
Note that in the string theory description, similar subtleties
may be encountered due to the way fundamental matter
is engineered geometrically via singular Calabi-Yau
compactification, or D-brane wrappings. 

Thus we are forced to take a different approach.  If one assumes that
the Lagrangian of the field theory is such that the graviphoton does
not couple to the fields in a way that modifies the propagator of the
chiral superfields, the techniques employed in \cite{Dijkgraaf:2002xd}
can still be used in the absence of gravity. Under such assumptions the
graviphoton dependence will show up as a result of applying the chiral
ring relations.  The chiral superfield propagator is given by
\cite{Dijkgraaf:2002xd}
\begin{equation}
\frac{1}{p^2 + W^\alpha \pi_\alpha + m} = \int_0^\infty ds e^{-s (p^2
+ W^\alpha \pi_\alpha +m)} \ ,
\end{equation}
where $p$ and $\pi_\alpha$ are the bosonic and fermionic momenta,
respectively, and using holomorphicity $\bar m$ has been taken to be
$1$. The vertices of the theory are read directly from the tree-level
superpotential. The computation then proceeds by integrating over both
the bosonic and fermionic loop-momenta. The bosonic integral is a
simple Gaussian integral yielding a determinant depending on the
Schwinger parameters $s_I$ while the fermionic one brings down
insertions of $W_\alpha$ which combine using the chiral ring relations
to form the graviphoton and glueball dependent terms multiplying a
polynomial of the Schwinger parameters. According to the conjecture
the fermionic and bosonic $s$-dependence should cancel leaving only a
vector model computation. The $W_\alpha$ insertions must be
path-ordered as these are no longer anti-commutative.

However, as we will soon demonstrate this approach fails. An example
of this is one of the graphs for the correlation function \ev{M}
\\ \centerline{\includegraphics{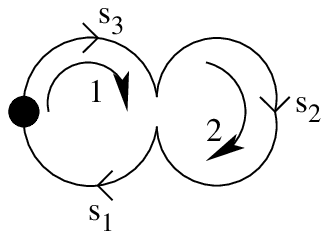}} \\
The bosonic integral is found to be
\begin{equation}
Z_B = \frac{1}{(4 \pi)^4 (s_1 + s_3)^2 s_2^2} \ ,
\end{equation}
while the fermionic integration with the origin of path-ordering taken
to be the $M$ insertion yields by utilizing the chiral ring relations
\begin{eqnarray}
Z_F & = & \frac{1}{16} s_2^2 \left[ (s_3^2 + s_1^2) \Tr (W^2 W^2) + 2
s_1 s_3 \Tr (W^\alpha W^2 W_\alpha) \right] = \nonumber \\
& = & {} -\frac{N_c}{32} s_2^2 (s_1 -s_3)^2 F_{\alpha \beta} F^{\alpha
\beta} \ ,
\end{eqnarray}
whose $s$-dependence does not cancel with that of $Z_B$.  But taking the
path-ordering origin at the vertex one has
\begin{equation}
Z_F = -\frac{N_c}{32} s_2^2 (s_1 + s_3)^2 F_{\alpha \beta} F^{\alpha
\beta}
\end{equation}
leading to the exact cancellation of the $s$-dependence of this
graph. One is drawn to conclude that this diagram should be taken to
be zero by some selection rule in order to remove this ambiguity.

Other graphs also feature such problems. For example, the
correlation function \ev{M^2} includes contribution from the graph \\
\centerline{\includegraphics{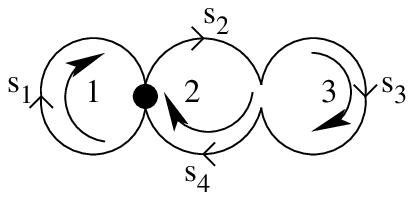}} \\
The bosonic integral in this case is
\begin{equation}
Z_B = \frac{1}{(4 \pi)^6 s_1^2 (s_2 + s_4)^2 s_3^2} \ ,
\end{equation}
while the fermionic integration with the origin of the path-ordering
taken to be at the $M^2$ insertion is
\begin{equation}
Z_F = \frac{\pi^2}{4} s_1^2 s_3^2 (s_2 - s_4)^2 F_{\alpha \beta} F^{\alpha
\beta} S \ .
\end{equation}
Taking the origin of the path-ordering at the other vertex yields the
same result. It can be seen that again in this case the $s$-dependence
does not cancel.

We conclude that either the field theory Lagrangian must also be
deformed in some way in addition to the $C$-deformation, or an
appropriate scheme has to be developed within the framework of the
above assumptions.

\acknowledgments
We would like to thank P.~A.~Grassi, V.~Kaplunovsky, K.~S.~Narain and
H.~Ooguri for valuable discussions.  This work is supported by the ISF
and the GIF.

\bibliographystyle{JHEP-2}
\bibliography{PhD}
\end{document}